\documentclass[lnicst,sechang,a4paper,pdftex]{svmultln}
\usepackage{amssymb}
\usepackage{graphicx}
\usepackage[caption=false]{subfig}

\DeclareGraphicsExtensions{.pdf,.png,.jpg,.eps}

\usepackage{url}
\urldef{\mailsa}\path|{first name. last name}@insa-lyon.fr|
\usepackage[pdfpagelabels,hypertexnames=false,breaklinks=true,bookmarksopen=true,bookmarksopenlevel=2]{hyperref}

\begin{document}

\mainmatter  

\title{Spontaneous Proximity Clouds:\\
 Making Mobile Devices to Collaborate for Resource and Data Sharing}

\titlerunning{ }

%
%
\author{Roya Golchay%
\and Fr\' ed\' eric Le Mou\"el\and Julien Ponge\and Nicolas Stouls}  %
\authorrunning{ }

\institute{University of Lyon, INSA-Lyon, INRIA CITI Lab,\\
F-69621 Villeurbanne, France\\
\mailsa
}

%
%

\toctitle{Lecture Notes in Business Information Processing}
\tocauthor{ }
\maketitle

\begin{abstract}

The base motivation of Mobile Cloud Computing was empowering mobile devices by application offloading onto powerful cloud resources. However, this goal can't entirely be reached because of the high offloading cost imposed by the long physical distance between the mobile device and the cloud. To address this issue, we propose an application offloading onto a nearby mobile cloud composed of the mobile devices in the vicinity - a Spontaneous Proximity Cloud. We introduce our proposed dynamic, ant-inspired, bi-objective offloading middleware - ACOMMA, and explain its extension to perform a close mobile application offloading. With the learning-based offloading decision-making process of ACOMMA, combined to the collaborative resource sharing, the mobile devices can cooperate for decision cache sharing. We evaluate the performance of ACOMMA in collaborative mode with real benchmarks - Face Recognition and Monte-Carlo algorithms - and achieve 50\% execution time gain. 
\end{abstract}

\section{Introduction}

Mobile Cloud Computing is the emerging paradigm of recent decades that focuses on overcoming the inherent shortages of mobile devices regarding processing power, memory and battery via application offloading, by total or partial execution of mobile applications on a distant cloud. Hence, the application offloading might not always be helpful because of the long physical distance between the mobile device and the cloud. The concept of the cloudlet~\cite{Mahadev2009} has been raised to response to this issue of distance. The cloudlet is a predefined cloud in proximity that consists of some static stations and is generally installed in public domains, but with no guaranty of availability near a mobile device.

As a solution to this cloud distance problem, we propose to offload application onto a Spontaneous Proximity Cloud~(SPC) - a cloud in the proximity of the mobile device, composed of a set of mobile devices in the vicinity. This SPC is a collaborative group of moving devices in proximity with members that occasionally join and leave. 

The short distance between the mobile device and the SPC overcomes the issue of latency in data transfer to distant clouds, especially in high network traffic conditions. Offloading onto SPC could also prevent imposing bandwidth allocation overhead onto a communication network that experiences a shortage of capacities, due to the continuous traffic growth. Besides, the energy consumption of a 3G cellular data interface, associated with the cloud, is 3 to 5 times much higher than WiFi transmissions, used between mobile devices~\cite{Cuervo2010}\cite{Miluzzo2012}. 
Another motivating factor to use SPC is the popularity of mobile devices. Inadequate network coverage, natural or man-made disasters may damage the data centres and significant technical failures - such as experimented by Amazon cloud~\cite{Adem2015} - can make remote clouds temporarily unavailable. While, because of the increasing number of mobile devices and the wide frequency of use - per user or household \cite{ahonen2011} - a mobile device presents a great chance to be surrounded by a group of mobile devices.
Finally, the use of SPC is a perfect incentive for green computing, with individual devices powered under the user responsibility that can use human body kinetic energy harvesting or solar panels~\cite{Adem2015}. 

We found all these factors motivating enough to design and implement ACOMMA, an \textbf{A}nt-inspired \textbf{C}ollaborative \textbf{O}ffloading \textbf{M}iddleware for \textbf{M}obile \textbf{A}pplications, that performs offloading on either distant cloud or SPC.
ACOMMA is an automated offloading middleware that takes offloading decisions dynamically by applying an ant-inspired bi-objective decision-making algorithm. The details of offloading onto distant cloud are already explained and evaluated in our previous article~\cite{Golchay2016}. In this paper, we demonstrate that taking a decision in a mobile device can benefit to all the other mobile devices in the vicinity and so better the application execution performances. We create a decision cache composed of the execution trails of mobile applications and, by using learning-based decision-making algorithm, ACOMMA could reuse previous offloading decisions instead of running its Ant Colony Optimization (ACO) decision-making algorithm. 

In this paper, we focus on the extension of ACOMMA in a way that it can be able to perform offloading in a collaborative manner. In collaborative offloading, instead of communicating with a distant cloud, the mobile device cooperates with SPC's members, for either resource or data sharing. Our main contributions consist of:
\begin{itemize}
\item Developing a decision-making process performing multi-destination offloading. To this end, we need to modify the ACO algorithm to take potential offloading decisions to remote clouds as well as mobile devices in the SPC, without any lock-in considerations to the number of devices. In this case, the mobile devices collaborate for resource sharing.
\item Developing a learned-based decision-making process to use the collaborative decision cache instead of the local cache. In this case, the mobile devices collaborate for data sharing. They share their local caches to create a richer collaborative cache that permits more efficient and relevant offloading decisions.
\end{itemize} 

The remainder of the paper is structured as follows: Section 2 discusses the existing offloading approaches. Section 3 explains the architecture of our proposed offloading middleware - ACOMMA. Section 4  and 5 show how ACOMMA is enhanced to make mobile devices collaborate for resource and cache sharing. Section 6 evaluates our offloading middleware under a range of scenarios and using different benchmarks. Finally, Section 7 provides a summary, conclusion and outline of future work. 

\section{Related work}

Recently, delegating total or partial application execution to more powerful machines instead of local devices - known as application offloading - has attracted attention to overcome resource limitations and to save the battery of mobile devices. 
A significant amount of researches has been performed in this domain to propose solutions to bring the cloud to the vicinity of the mobile device 

MAUI~\cite{Cuervo2010} and ThinkAir~\cite{Kosta2012} are the most prominent works in this domain. They are focusing on optimising energy consumption or execution time using linear programming. They use the virtual machine migration techniques to execute application methods onto the cloud. However, these virtualized environments are heavy for limited mobile devices.
CloneCloud\cite{Chun2011} is a lighter approach since it cuts the application into two thread level partitions using linear programming, with only one of them offloaded onto the cloud. 
Some approaches perform offloading onto a closer surrogate, a cloudlet \cite{Mahadev2009}, that is composed of static stations. However,  a cloudlet does not necessarily exist near a mobile device. 

Few studies focused on the use of adjacent mobile devices as offloading surrogates. Transient cloud\cite{Penner2014} uses the collective capabilities of nearby devices in an ad-hoc network to meet the needs of the mobile device. A modified Hungarian method is applied as an assignment algorithm to assign tasks to devices that are to be run according to their abilities. The execution of each task by any device imposes some cost, and the assignment algorithm aims to find the minimum total cost assignment. To that end, \cite{Penner2014} has proposed a dynamic cost adjustment to balance the tasks based on costs between devices. 
Miluzzo et al. \cite{Miluzzo2012} proposed an architecture named mCloud that runs resource-intensive applications on collections of cooperating mobile devices and discuss its advantages. Kassahun et al. \cite{Adem2015} have gone one step further and formulated a decision algorithm for global adaptive offloading. They implemented the program components on mobile devices set to optimise Time to Failure (TTF) while taking into account the limitations of the effectiveness of the program. Having highlighted the benefits of collaboration for mobile task offloading, Mtibaa et al. also implemented computational offloading schemes to maximise the longevity of mobile devices\cite{Mtibaa2014}\cite{Mtibaa2013}.

\section{The general architecture of ACOMMA}
\begin{figure}
\centering
\includegraphics[height=8cm]{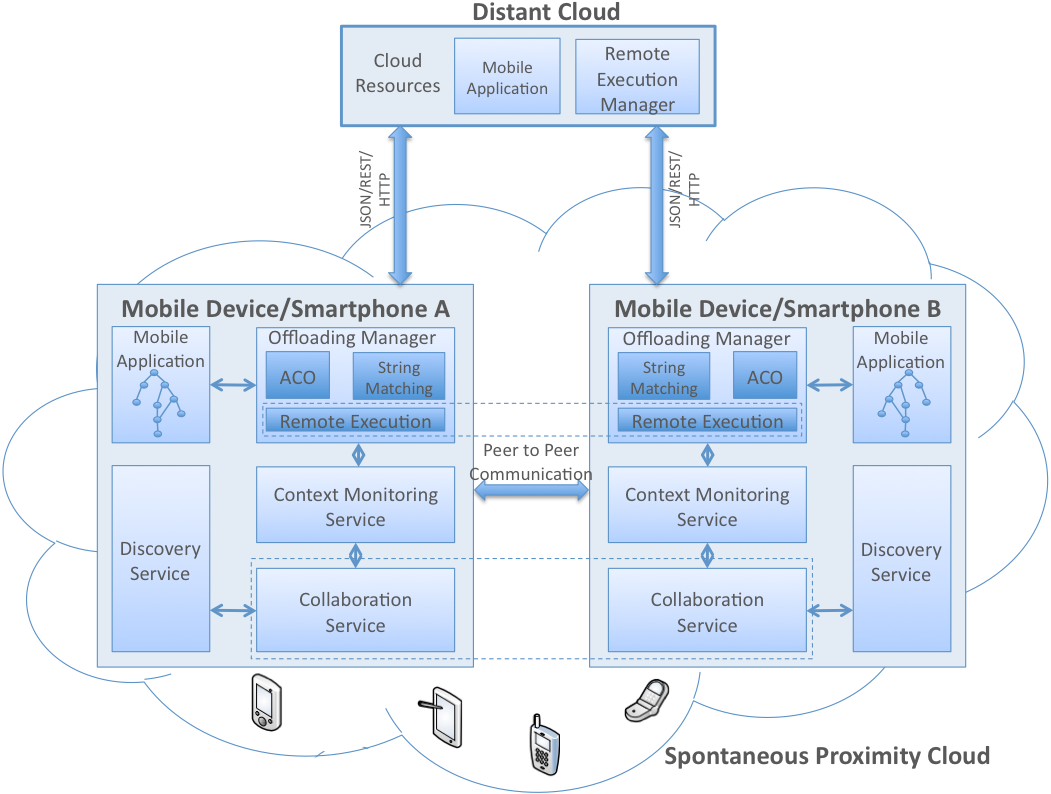}
\caption{The general architecture and building blocks of ACOMMA}
\label{fig:architecture}
\end{figure}

The proposed architecture of ACOMMA makes application offloading possible onto remote clouds and SPC as a single or multiple destination offloading process. The building blocks of ACOMMA are illustrated in Figure~\ref{fig:architecture}. 

ACOMMA considers a mobile application as a dependency graph, where the nodes represent the function/method calls of the application and the edges are their dependency in terms of function/method invocations.  The offloading decision-making process partitions this call graph to define which function/method should be executed locally - on the mobile device, near-remotely - on a device of the SPC, or far-remotely - on the distant cloud.

The offloading middleware is composed of a group of services to offload this application.
The \textit{Offloading Manager} is in charge of taking offloading decisions using (1) an Ant Colony Optimization algorithm for the initial decision-making or (2) String matching algorithm for learning-based decision-making. In the learning-based mode, the decision-making relies on previous application execution traces, saved in a local or collaborative decision cache. Coming to the collaborative mode, the \textit{Collaboration Service} takes the responsibility of offloading onto SPC with the help of \textit{Offloading Manager}. The \textit{Collaboration Service} makes nearby devices collaborate using the neighbours' information prepared by the \textit{Discovery Service}. This service finds the nearby devices and saves their address and information. 
To perform a dynamic offloading considering the current state of mobile devices, ACOMMA needs to be aware of current conditions and requirements. The mobile devices' information, such as the available battery and memory, and their environment such as the available networks, the available bandwidth, as well as cloud kind and theirs costs, are collected by \textit{Context Monitoring Service}. This contextual information helps ACOMMA to choose in-between the SPC or the remote clouds.



\section{Collaborative resource sharing in application offloading}
\begin{figure}
\centering
\includegraphics[height=4cm]{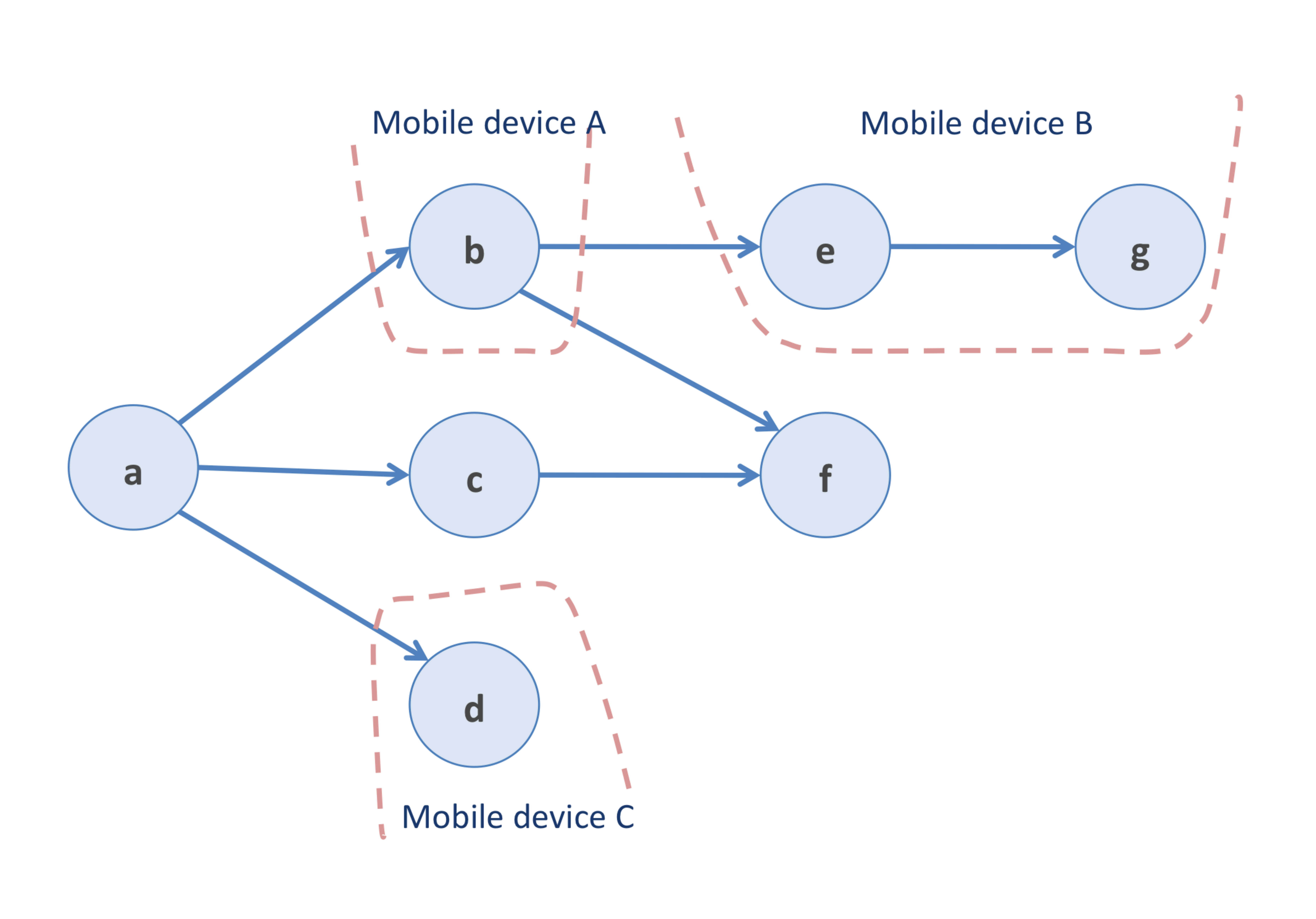}
\caption{Application partitioning for multi destination offloading}
\label{fig:partitioning}
\end{figure}

As mentioned before, the decision-making process of ACOMMA is based on the application call graph partitioning. To perform offloading onto SPC, the decision engine breaks apart the application into several parts - instead of two in traditional partitioning approaches, where each part represents an executing device. For example, Figure~\ref{fig:partitioning} shows the partitioning for offloading onto three nearby devices, where nodes a, c, f execute locally, node b executes on device A, node e and g execute on device B and finally mobile device C executes node d.

\begin{figure}
\centering
\includegraphics[height=4cm]{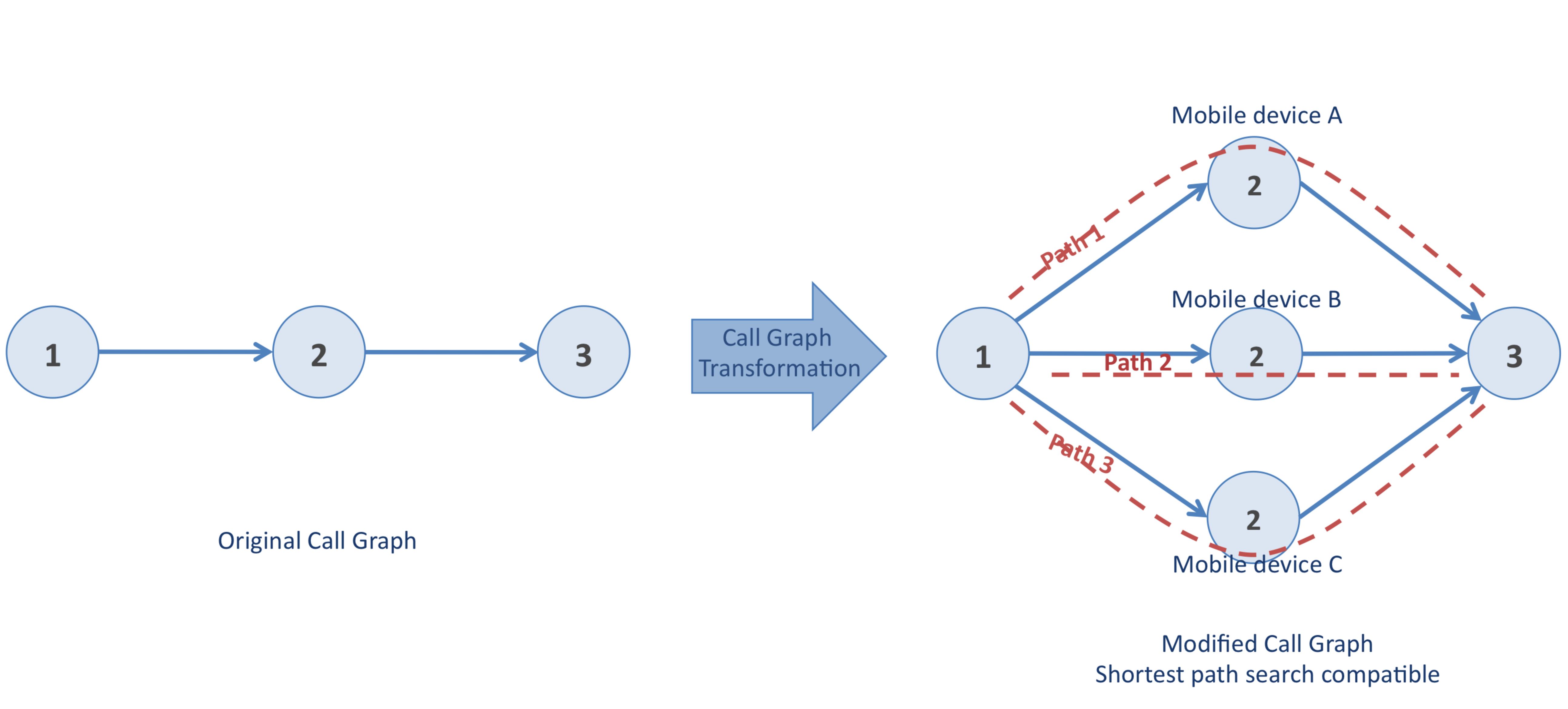}
\caption{Call graph modification for multi destination offloading}
\label{fig:transformation}
\end{figure}

To perform such a graph partitioning for multi-destination offloading, the ACOMMA collaboration service modifies the application call graph in a way that for each method, several nodes are added to the graph, depending on the number of potential executing devices, one for each device. The modification process of the call graph is shown in the Figure~\ref{fig:transformation}. 

The original graph is composed of three nodes, where the start and end nodes (node 1 and 3) have to execute locally. Assuming that there are two devices in SPC in addition to the current mobile device, the node 2 is then duplicated two times - as often as the number of possible execution targets. The ACOMMA decision engine partitions the graph using an ACO algorithm that finds the shortest path between the start and end points of the graph. The choice of the first path shows the local execution of method 2 on device A, where the choice of the second and the third path represents the execution of method 2 on device B and device C respectively. 

Finding shortest path is done according to weights assigned to the edges of the graph. Since the different devices can have different optimisation goals, to reach a consensus in the objective function, we apply a multi-objective decision-making process - illustrated by a bi-objective decision-making with the execution cost of the related method regarding CPU usage and execution time.To take dynamic offloading decisions based on the current state, the shortest path is calculated for each function/method call in the total graph. 

\section{Collaborative decision cache sharing}
Learning is one of the primary functions of dynamic systems - such as in sensor networks and mobile networks. It is mainly used for the establishment of a relevant situation and the adaptations to the environment. In existing SPC, the learning process stays local. We argue that, when a mobile device takes a decision, this decision could benefit to the other devices nearby. 

To distribute the local decisions, we rely on a sharing decision cache. The sharing decision cache between nearby devices makes collaborative decisions possible. In this learning-based decision-making process, the mobile devices in the same state and environmental conditions could perform offloading in the same way as their neighbours. Moreover, even if the execution conditions are not exactly the same, in case of common applications, the decision is relevant enough. 

To take collaborative decisions, the collaborative cache is created by merging local cache of nearby devices. They could receive and send respecting different dissemination, merging and invalidating policies. 
For receiving neighbour's local decisions, we propose on-demand, periodical and on-change policies. Using an on-demand method, a mobile device broadcasts a cache request to the nearby devices whenever needed. In the periodic method, each mobile device periodically sends their decision caches to their neighbours without any concerns about their requirement. Also, in the on-change method, the source device sends its decision cache whenever it is modified either by adding a new execution trail or deleting old ones. 
The merge could be done simply by adding the new executing trail at the end of the local cache. Alternatively, another way is to implement a collaborative cache with unique rows by deleting the duplicate traces. Creating a weighted cache is also an implementation available. The weight of each executing trace corresponds to the number of decisions already taken - implying that an already optimisation decision have more chance to be reselected.
As cache invalidation policies, we propose periodic and on-change methods. While the offloading decisions highly depend on the current status of the mobile device itself and its environment, the cache could reset when these conditions change.

Applying different combinations of these policies for cache management may greatly impact the performance of ACOMMA for offloading using collaborative decision-making.

\section{Implementation and evaluations}
\subsection{Benchmark applications and experimental platform}
To evaluate the performance of ACOMMA, we first test micro-benchmarks mathematical functions: Matrix determinant and Integral - consuming enough resources to make offloading valuable. Their small number of methods helps us to trace function call executions accurately. To be closer to real applications, we also implement macro-benchmarks with popular offloading applications of Chess and Face Recognition including Monte Carlo and Eigenfaces algorithms~\cite{face}.

As mobile devices, we use Samsung Galaxy SII with 1,2 GHz dual-core processor and 2GB of memory running Android version 4.1.2 (Jelly Bean) and Asus Google Nexus 7 pad with quad-core 1.2 GHz processor and 1GB of memory running Android version 5.1.1 (Lollipop). To successfully validate collaborative offloading of ACOMMA using SPC, we need to show that ACOMMA can detect the SPC and correctly dispatch the methods of running application between detected nearby devices according to their processing power. 

\subsection{Results}
To evaluate the efficiency of offloading onto the SPC, we apply a scenario where a Galaxy SII makes offloading onto an SPC that consists in 2 Galaxy SII and 3 Google Nexus 7 pads. We compare the performance of offloading onto this SPC with offloading onto a MacBook Pro with 8 GB of memory, a 250 GB hard disk and a 2,53 GHz Intel processor dual-core as a remote server. This server has OS X 10.9.5 Mavericks as operating system. The result shows that the local execution is rather slow - 1200ms for the Monte Carlo application, and offloading onto the MacBook presents a significant gain in terms of execution time - 60-70ms~\cite{Golchay2016}. Offloading to the SPC - less powerful than a remote cloud, but with a better latency - results in a less efficient execution time - 100ms, but interesting enough to test the benefit of a collaborative cache.

Coming to the evaluation of dispatching onto SPC, we ran Determinant and Integral 10 times with a SPC composed of four devices in addition to the source device. In this scenario, D1, D2, D3 - Google Nexus pads - and D4 - Samsung Galaxy SII - are offloading destination devices.
\begin{figure}
\centering
\subfloat[Integral offloading]{\includegraphics[height=4cm]{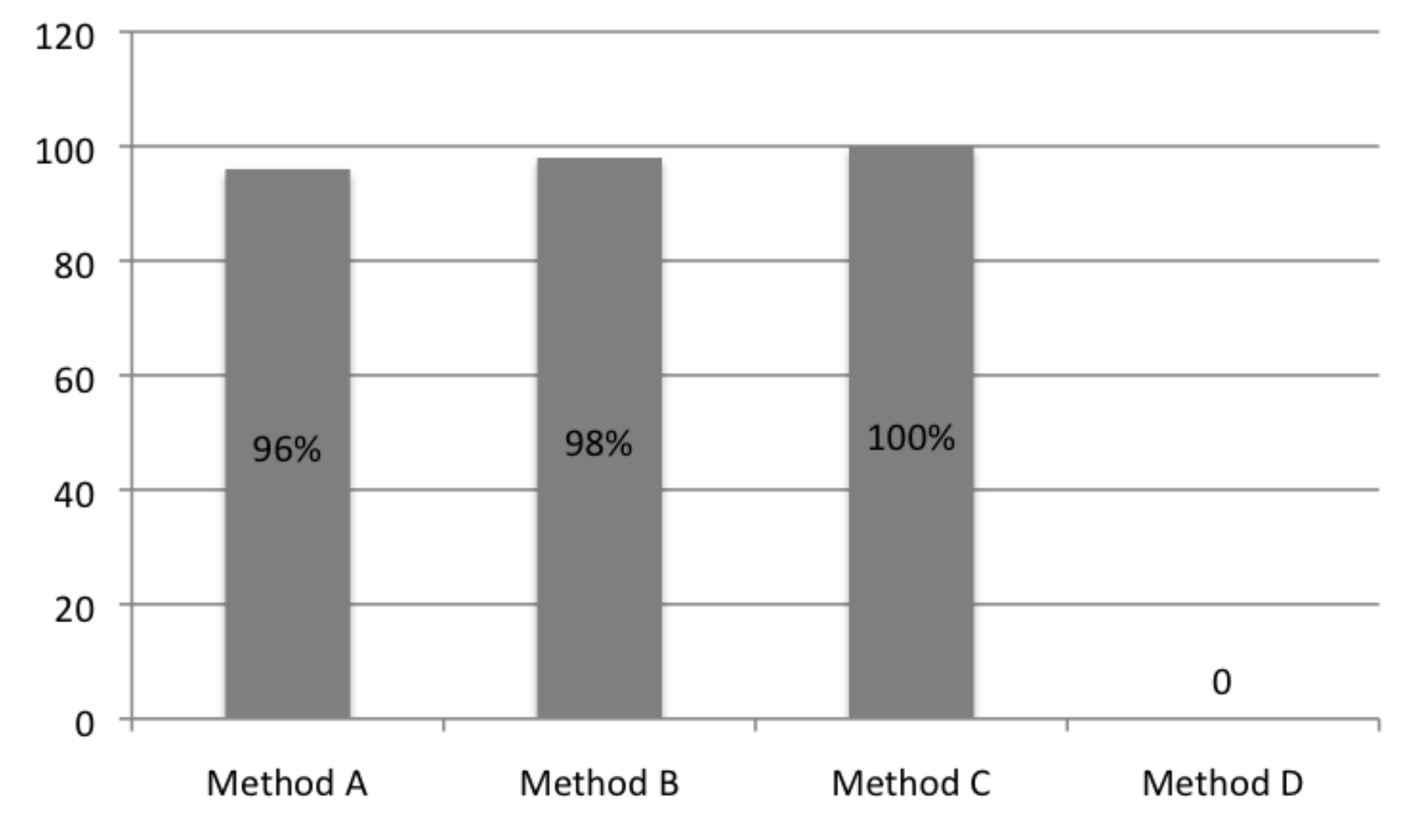}} 
\subfloat[Determinant offloading]{\includegraphics[height=4cm]{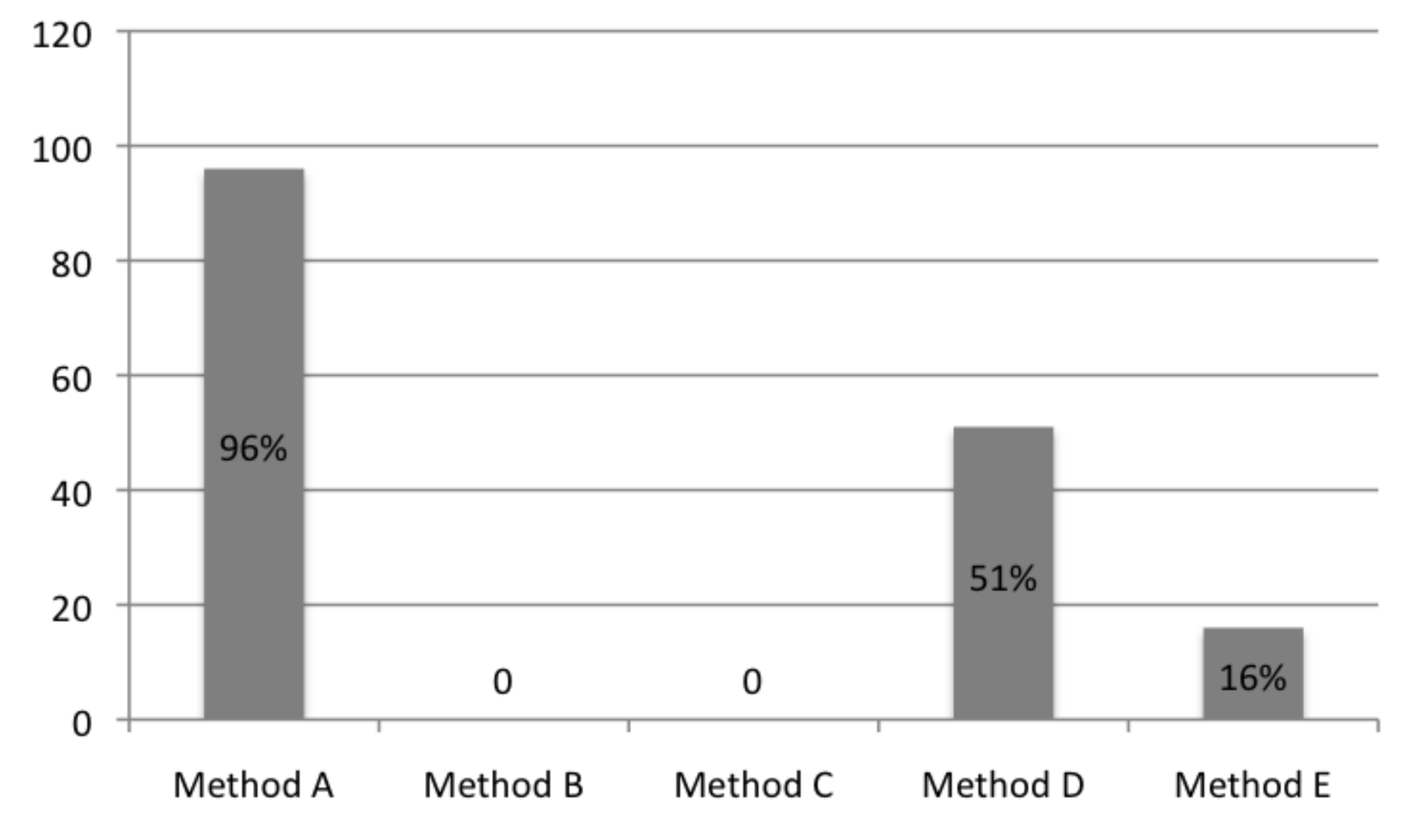}}
\caption{Successful method offloading rate(\%)} 
\label{fig:offloading} 
\end{figure} 

Figures ~\ref{fig:offloading}(a) and ~\ref{fig:offloading}(b) show the percentage of successful offloading for each application method. Considering the four methods of Integral benchmark, the method D is never offloaded. Considering the five methods of Determinant, method B and method C were always executed locally. These are the methods that consume a negligible amount of resources, when offloading them impose more cost to the system compared to their local execution.

\begin{figure}
\centering
\subfloat[Integral]{\includegraphics[height=4cm]{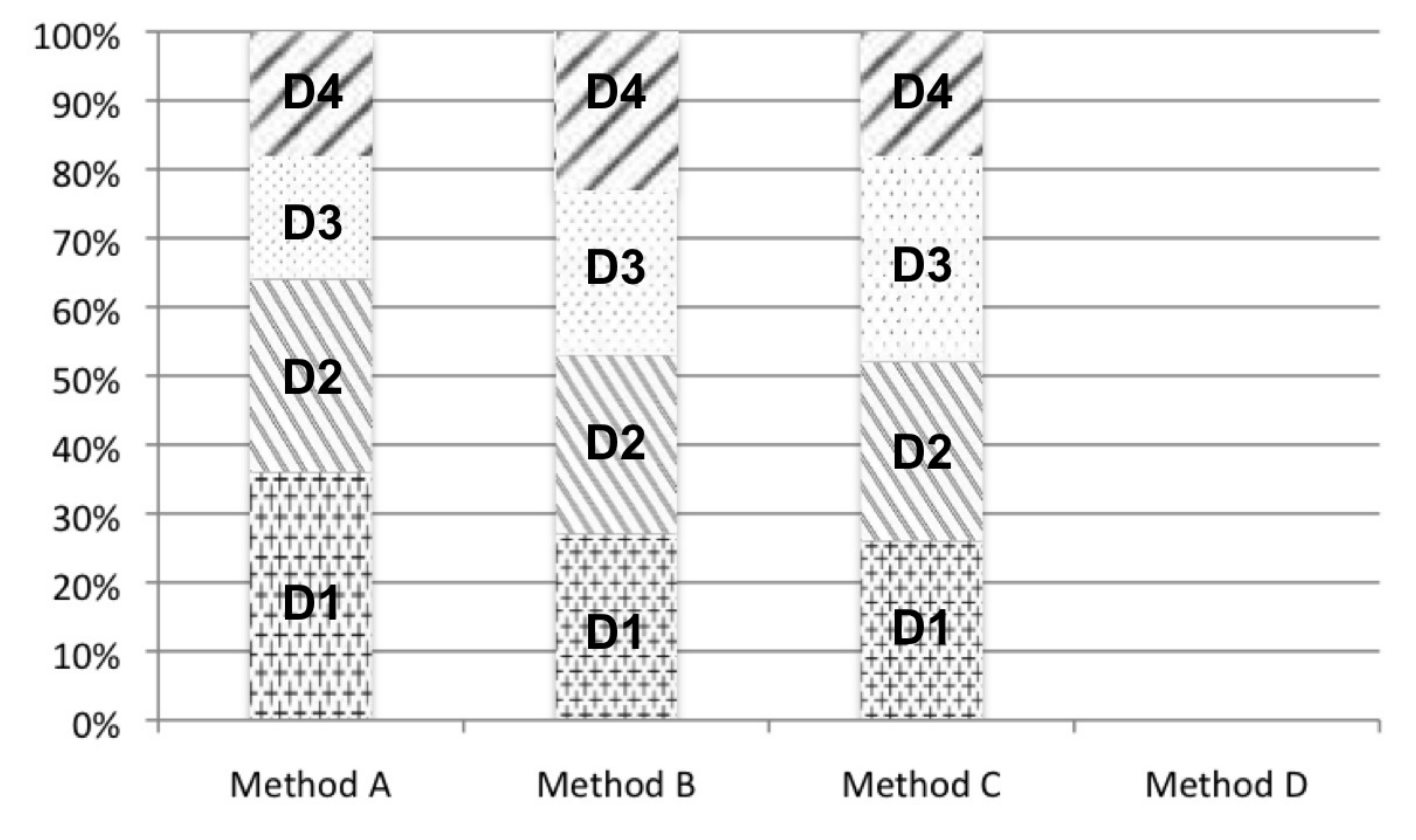}} 
\subfloat[Determinant]{\includegraphics[height=4cm]{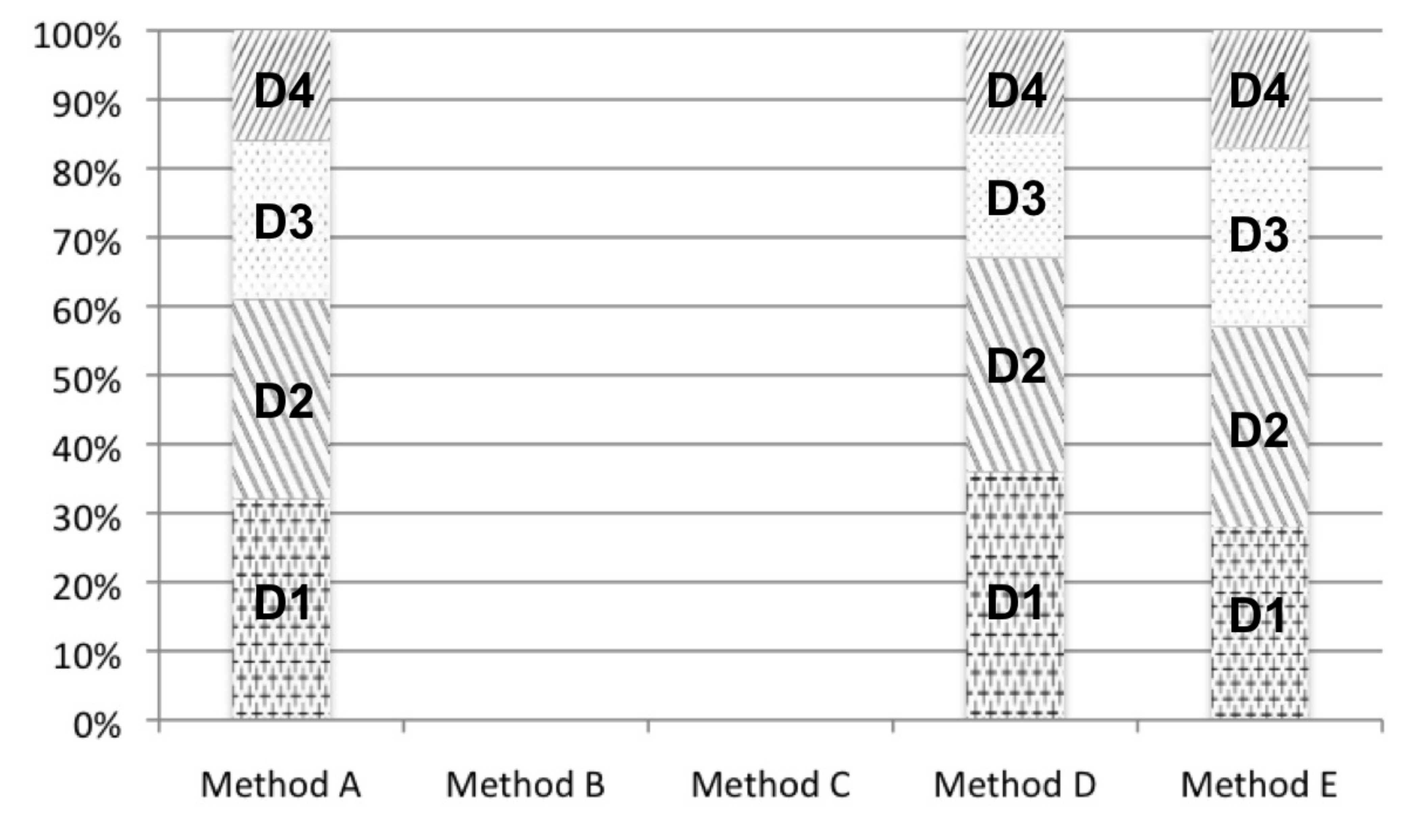}}
\caption{Contribution of devices to execute offloaded methods(\%)} 
\label{fig:dispatch} 
\end{figure} 

Figures ~\ref{fig:dispatch}(a) and ~\ref{fig:dispatch}(b) shows the offloading proportion and the contribution of each device to the offloading process. As expected, in most of the cases the device D4 has the lighter portion of execution as it is the less powerful device. The devices D1, D2 and D3 that have the same hardware characteristics, have almost the same execution contribution, even if their loads are not exactly equal. These results show that ACOMMA is really context- and application-aware to make multi-destination offloading and to dispatch the application methods between SPC's members correctly.

To evaluate the sharing benefits, we compare the execution time of the different benchmarks between ACO and string matching based on a local cache~\cite{Golchay2016} and the collaborative cache. We evaluate the decision-making based on the collaborative cache with an on-changed dissemination policy, and a unique weighted cache merge policy (the invalidating policy has been shown to have a neglectable impact on results~\cite{Golchay2016}). We run Determinant, Integral, Face detection and Monte-Carlo 10 times for two series of inputs on Galaxy SII. First, the ACO algorithm offloads and populates the local cache with maximum 10 rows. Secondly, when the source device has finished its executions, we run the same application on destination devices while making the offloading decision using collaborative cache and a string matching algorithm.

\begin{figure}
\centering
\includegraphics[height=5cm]{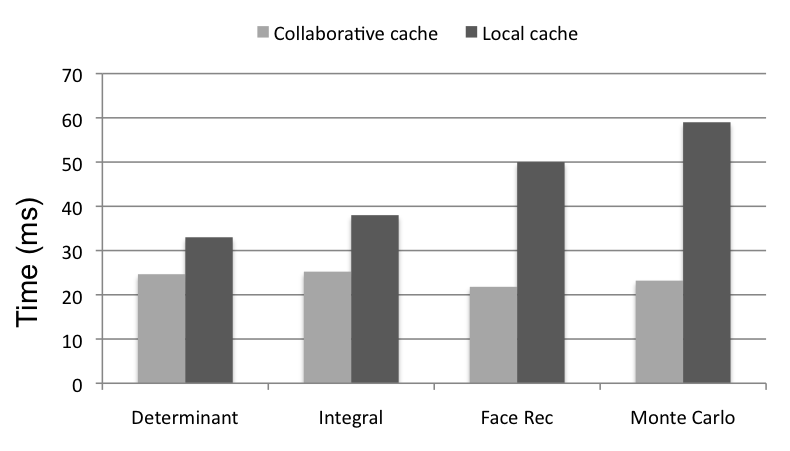}
\caption{Learned based offloading using local and collaborative cache}
\label{fig:collaborative-cache}
\end{figure}

The results in Figure~\ref{fig:collaborative-cache} show that learning-based decision-making using a collaborative cache is more efficient than using a local cache. The gain in terms of execution time depends on the graph size. Small applications - Determinant and Integral - presents 30-40\% gain (10ms) and more complex ones - Face Recognition and Monte Carlo - up to 60\% (35ms).

\section{Conclusion and future work}
In this work, we propose the Spontaneous Proximity Cloud concept to offload applications onto mobile devices in the vicinity. We enriched our ant-inspired bi-objective offloading middleware, ACOMMA, with a learning-based decision-making using a collaborative resource and data cache sharing. We evaluate the performance of ACOMMA in collaborative mode with real benchmarks - Face Recognition and Monte-Carlo algorithms - and achieve 50\% execution time gain. 
Several issues need further work. Testing the robustness and the scalability of the offloading to the mobility and connection interferences is a major one. A balance between the caching cost - storage and network - and the better decision to take has to be carefully studied.



\end{document}